\documentstyle[11pt,paspconf,epsf]{article}

\begin{document}

\title{Galaxies and Large Scale Structure at $z \sim 3$}

\author{C. Steidel\altaffilmark{1}, K. Adelberger\altaffilmark{1}, M. Giavalisco\altaffilmark{2},
M. Dickinson\altaffilmark{3}, M. Pettini\altaffilmark{4}, and M. Kellogg\altaffilmark{1}}


\altaffiltext{1}{Palomar Observatory, Caltech 105-24, Pasadena, CA 91125, USA}
\altaffiltext{2}{Carnegie Observatories, 813 Santa Barbara St., Pasadena, CA 91107, USA}
\altaffiltext{3}{The Johns Hopkins University, and STScI, 3700 San Martin Drive, Baltimore, MD 21218}
\altaffiltext{4}{Royal Greenwich Observatory, Madingley Road, Cambridge CB3 OEZ, UK}



\begin{abstract}
We summarize the status of a ``targeted'' redshift
survey aimed at establishing the properties of galaxies and
their large scale distribution in the redshift range $2.5 < z < 3.5$.
At the time of this writing, we have obtained spectra of more than
400 galaxies in in this redshift range, all identified using
the ``Lyman break'' color--selection technique.  We present some of
the first results on the general clustering properties of the Lyman
break galaxies.
The galaxies are very strongly clustered, with
co-moving correlation length similar to present--day galaxies, and
they are evidently strongly biased relative to the mass distribution
at these early epochs, which is consistent with hierarchical galaxy
formation models if Lyman break galaxies trace the most massive
halos at $z \sim 3$. 
Prospects for large surveys for galaxies beyond $z \sim 4$
are discussed. 

\end{abstract}


\keywords{globular clusters,peanut clusters,bosons,bozos}


\section{Introduction}

At the time of this writing (December 1997), it has been slightly more than 2 years since
the ``Lyman break'' technique (discussed extensively elsewhere, e.g. 
Steidel \& Hamilton 1993, Steidel, Pettini, \& Hamilton 1995, Madau  et al. 1996, etc.), 
for isolating large numbers of high redshift
($z \sim 3$) galaxies was demonstrated to work efficiently, based on
confirming spectra obtained at the W.M. Keck Observatory (Steidel et al. 1996).
Galaxies at $z\sim 3$ or greater are no longer terribly surprising, as now the highest
redshift galaxy has $z=4.92$ (Franx et al 1997) and it probably won't be very long before
the $z=5$ barrier is surpassed. However, as exciting as finding distant galaxies
can be, the greatest gains for our understanding of the galaxy formation process
will come from the feasibility of assembling large samples of galaxies
at previously inaccessible redshifts, and to study their properties and
distribution with a high level of statistical significance. In view of this,
we have spent the ensuing 2 years compiling a large sample of $z \sim 3$ Lyman
break galaxies, selected in a consistent manner over a number of relatively
large fields. Our aim in carrying out this survey is to obtain $\sim 100+$ redshifts
per field in each of 5--6 fields, each of size 160 square arc minutes or greater.
We anticipate that the present survey will reach completion by the end of 1998.  

Our survey for Lyman break galaxies continues to be based upon deep ground--based
images obtained in a custom broad--band filter system, $U_nG{\cal R}$. For reliable
identification of high redshift galaxy candidates, we have found that long integrations
on 4m--class telescopes are required to reach adequate depth and S/N; most of
our images have been obtained at the prime focus of the Palomar 5m telescope, and
a high quality data set for a 9\arcmin\ by 9 \arcmin\ field of view requires about
12 hours of total integration time (7 hours in the $U_n$ band alone), or about
two good nights of observing. The identification of candidate $z \sim 3$ galaxies
is based upon a combination of the modeling of the spectral energy distributions
of high redshift star--forming galaxies, including the effects of the opacity of
the galaxy ISM and the intergalactic medium (cf. Madau 1995), and the results
of our spectroscopy, which have resulted in a somewhat modified color--selection
window as compared to the purely theoretical one which was used originally
(cf. Steidel, Pettini, \& Hamilton 1995). Figure 1 shows a two--color diagram
from one pointing of the Palomar 5m, indicating the region from which
we select our samples of Lyman break galaxy candidates. Approximately
5\% of all galaxies to ${\cal R}=25.5$ are candidate high redshift ($2.5 < z < 3.5$) galaxies. 
\begin{figure}
\plotone{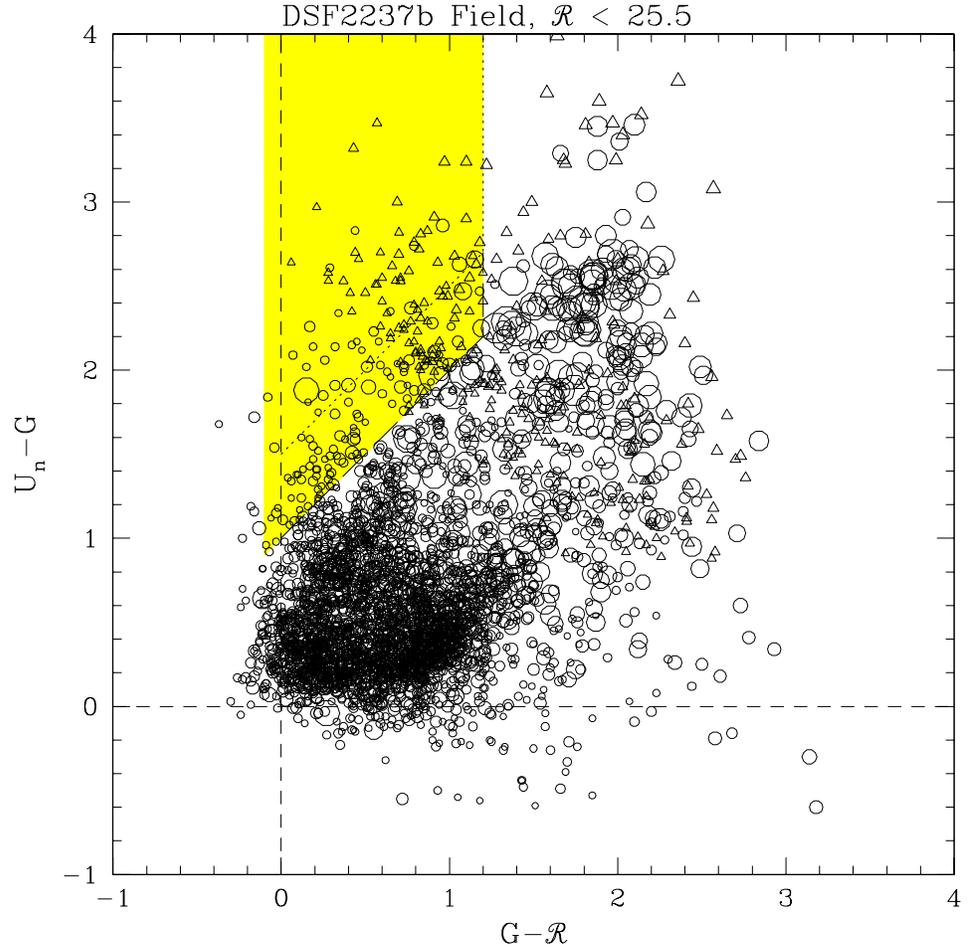}
\caption{An example of a two--color diagram used to select Lyman break
galaxies for follow--up spectroscopy. The shaded region corresponds
to the region from which candidate Lyman break galaxies are selected. 
Of the $\sim$3300 objects in the field
to ${\cal R}=25.5$, approximately 170 of them satisfy our adopted color criteria. 
Triangles represent objects for which the $U_n-G$ color is a lower limit.
 }
\end{figure}
\begin{figure}
\plotfiddle{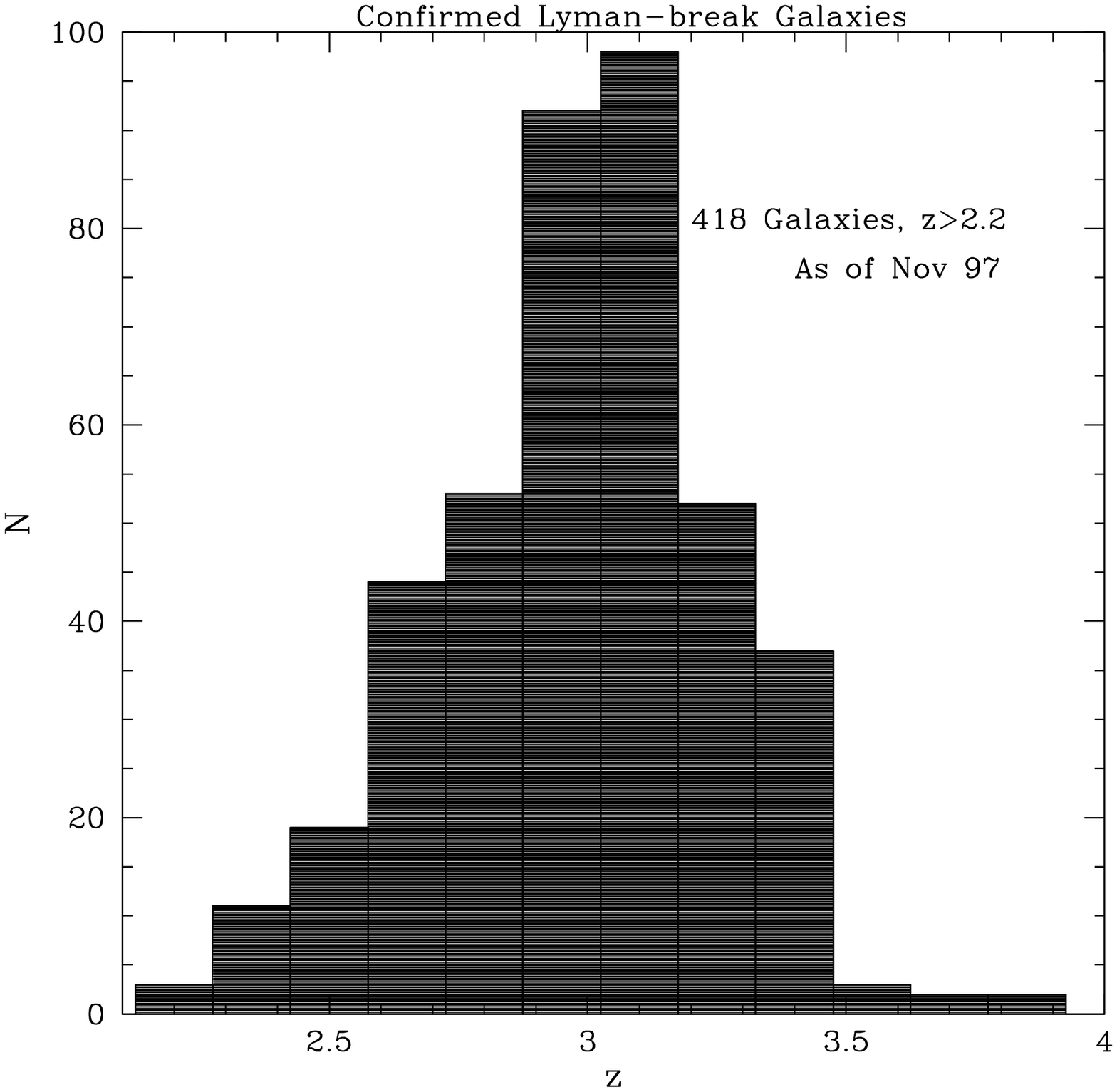}{2.5in}{0}{40}{40}{-120}{-80}  
\caption{Redshift histogram of spectroscopically confirmed Lyman break
galaxies in our $z \sim 3$ survey. The median redshift is $\langle z \rangle =
3.02$. The redshift cutoff near $z \sim 3.4$ is a result of our color selection
criterion in $G-{\cal R}$, while the lower redshift cutoff is imposed by
the requirement that a significant ``break'' must be present between the
$U_n$ and $G$ passbands (cf. Fig. 1). }
\end{figure}

\section{Spectroscopic Results}

All of the follow--up spectroscopy has been obtained using the Low Resolution Imaging
Spectrograph (LRIS) (Oke et al. 1995) on the W.M. Keck telescopes. On a typical slit
mask, we can obtain $\sim 20$ spectra of candidate $z \sim 3$ galaxies; our current
configuration generally leads to $\sim 80$\% of the observed galaxies yielding redshifts.
A clear night will yield a total of $\sim 50$ $z \sim 3$ galaxies. The current redshift
histogram for objects satisfying the color criteria in the shaded region of Figure 1
is shown in Figure 2; about 90\% of the galaxies fall in the redshift interval 
$2.6 \le z \le 3.4$, with none having $z < 2.2$. 

\begin{figure}
\plotfiddle{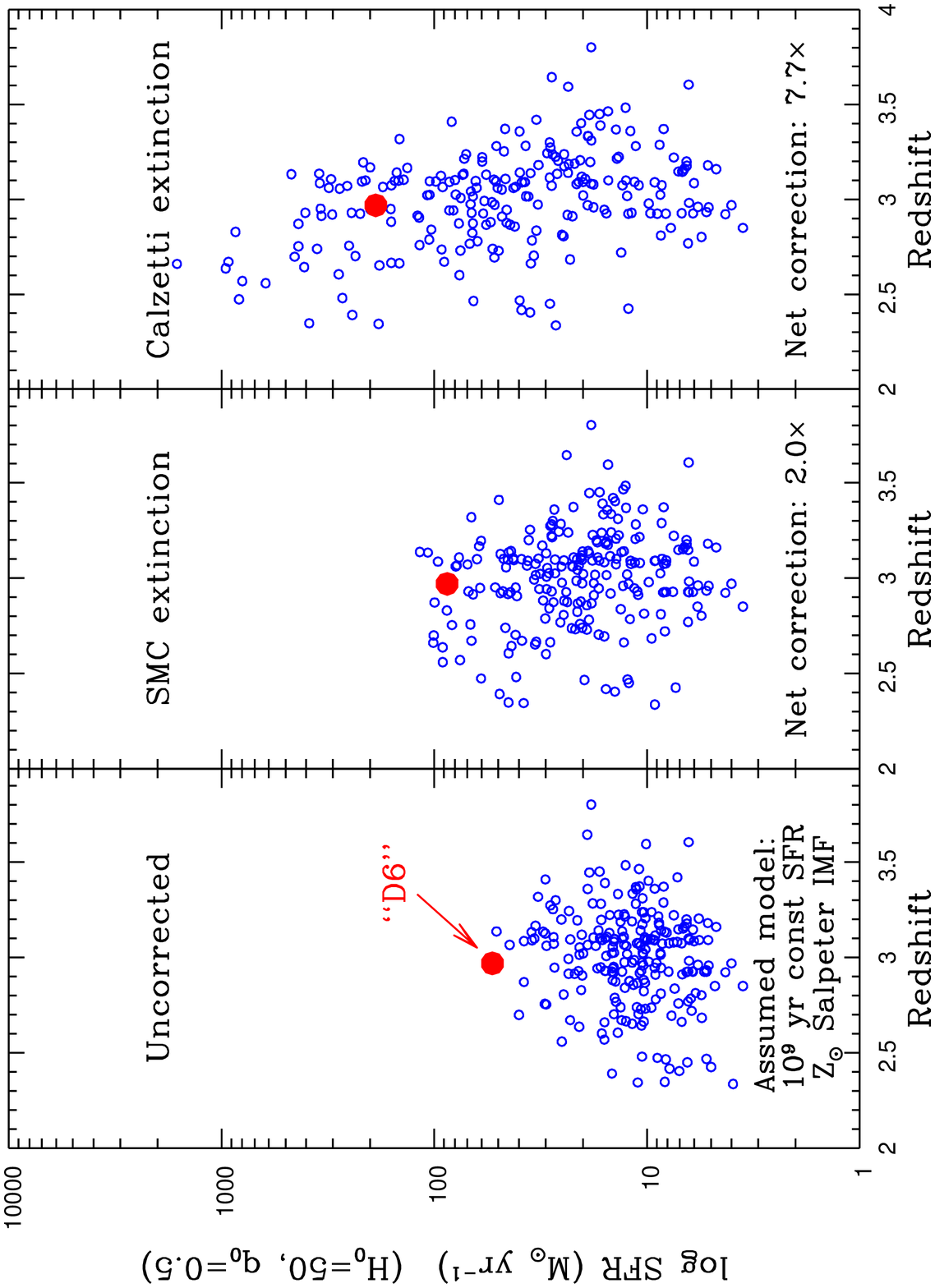}{3in}{-90}{40}{40}{-150}{250}
\caption{``De-reddening'' the observed Lyman break galaxies, and
the effects on the inferred star formation rates. The left panel
shows uncorrected UV luminosities translated into inferred star
formation rates; the middle and right panels show corrected versions
using two different reddening/extinction laws.}
\end{figure}

The success of the spectroscopic follow--up, and the well-established redshift
selection function for our particular color selection criteria, allow a number
of inferences based on the statistics of the larger photometric sample, which
at the time of this writing includes $\sim 1400$ Lyman break galaxy candidates
in a total survey area of about 0.25 square degrees. For comparison to our
earlier results prior to most of the spectroscopy, the surface density of
Lyman break galaxies having the redshift distribution shown in Figure 2, and brighter
than ${\cal R} =25.0$, is 1.0 arcmin$^{-2}$, or roughly 2.5 times larger than
our initial (conservative) estimates of the surface density of high redshift
galaxies. Clearly, these numbers are sensitive to the exact color cut adopted,
and the choice of limiting magnitude. 
We have converged on the color selection
region illustrated in Figure 1, and our photometry allows
reasonably complete identification of $z \sim 3$ Lyman break galaxies to
${\cal R}=25.5$.  
We will discuss details of the
redshift selection function, and the space density and luminosity distribution of the
LBGs, in Dickinson et al (1998).  

The combination of the photometry and the follow--up spectroscopy also enables
us to attempt an object--by--object extinction correction, 
by ``de-reddening'' each galaxy to the spectral energy distribution expected
for an unattenuated actively star--forming galaxy (further details and
results will be presented in Dickinson et al 1998) placed at the known
redshift. 
By far the largest uncertainty in doing this
is the adopted reddening/extinction model; as can be seen in Figure 3, the net
correction to the {\it observed} population of galaxies in terms of implied
star formation rate ranges from a factor of $\sim 2$ if an SMC--like reddening
curve is adopted, to a factor of $\sim 8$ for the assumption that the
empirical starburst galaxy extinction law of Calzetti (1997) is more appropriate.
We believe that these two reddening laws probably bracket the ``truth'', although
it will take much more work to establish the importance of extinction with
any certainty. For example, we are in the process of obtaining near--IR spectra (rest--frame optical)
of a sub-sample of Lyman break galaxies (Pettini et al 1998), which will provide a cross-check on the
appropriate extinction corrections using the luminosity of nebular emission lines.  
Rest--frame far--IR (observed sub-millimeter) observations of the same galaxies are
obviously of interest. 

\section{The Clustering of $z\sim 3$ Lyman Break Galaxies}

One of our main goals in undertaking a large survey of galaxies
at $z \sim 3$ was to investigate the large--scale distribution of
star--forming galaxies at such early epochs. Some of the preliminary
results of our spectroscopy have been published in Steidel et al. (1998), in which
the implications of a large structure of galaxies at $z=3.09$ were discussed.
The results of using the photometric sample, together with the well-established
redshift selection function, for an analysis of the angular clustering 
of $z \sim 3$ Lyman break galaxies are presented in Giavalisco et al (1998). 
A count--in--cells analysis of the survey fields in which we have
completed our spectroscopy will appear in Adelberger et al (1998).

The most striking (and robust) result at present is that the Lyman break
galaxies {\it must} be much more clustered than the overall mass distribution
on scales of $\sim 10h^{-1}$ Mpc, for any reasonable cosmological model.
Within the context of Cold Dark Matter models for the formation of structure,
this strong ``bias'' (we measure an effective linear bias parameter
of $b\approx 6$ for $\Omega_m=1$ and $b \approx 2$ for $\Omega_m=0.2$)
is expected if the LBGs are tracing relatively
massive dark matter halos, with $M_h \sim 10^{12}$ M$_{\sun}$ or greater. 
In hierarchical models, because halos with such large bias would be expected to
be tracing the densest environments at $z \sim 3$, the LBGs are probably
the progenitors of objects that end up in groups and clusters of galaxies
by the present day (Governato et al 1998, Wechsler et al 1998). 
As can be seen in Figure 4, large structures in redshift space are relatively common 
features, as we had asserted in Steidel et al. (1998); the largest peaks
are expected to become Coma--like clusters of galaxies by $z \sim 0$ (see the
article by Carlos Frenk in these proceedings).
\begin{figure}
\plotone{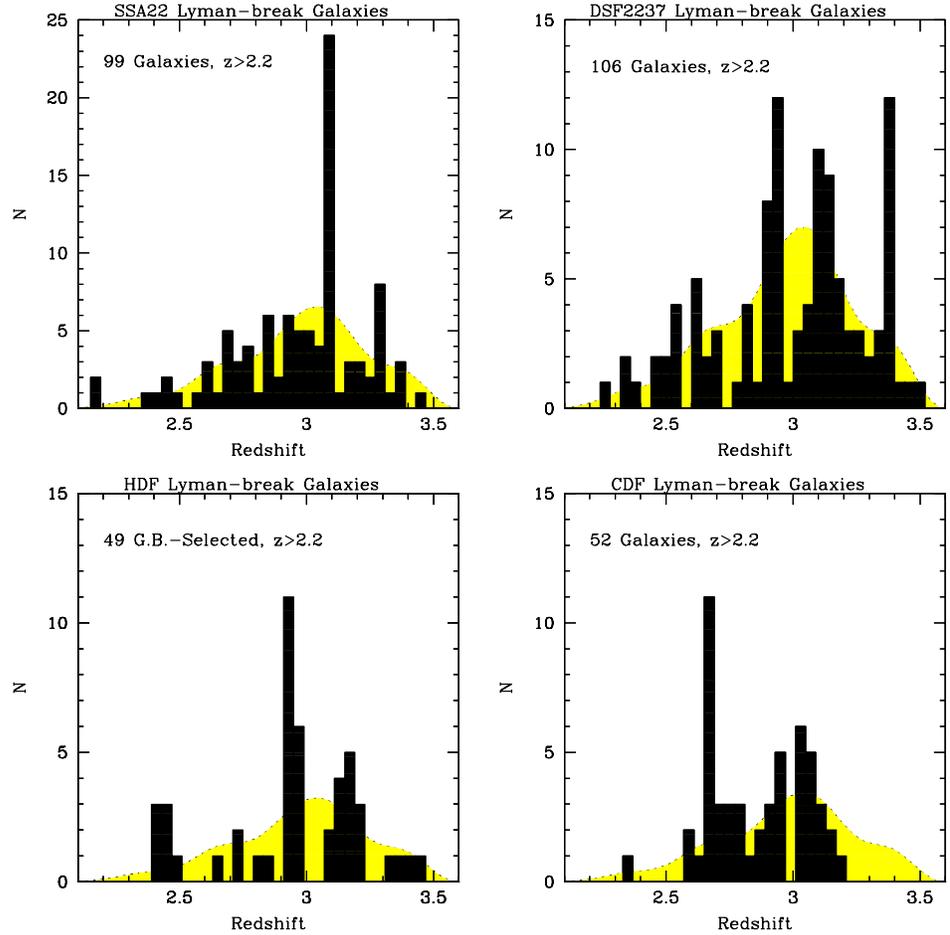}
\caption{Field--by--field redshift histograms from 4 of our survey fields.
DSF2237 and SSA22 are 9\arcmin\ by 18\arcmin\ fields, whereas
HDF and CDF are (at present) only single 9\arcmin\ by 9\arcmin\ fields. 
Note that there is at least one highly significant redshift--space
``structure'' per field. The most significant features are likely
to be concentrations that will become moderately rich clusters
by the present epoch. The variance in the redshift histograms
(relative to the selection function, which is shown with the dotted
curve in each panel) provides
a direct measure of the galaxy fluctuations on $\sim 10$ Mpc scales, which
can be compared with the fluctuations in the mass expected for various
assumed cosmologies.  
 }
\end{figure}
The co-moving correlation length for Lyman break galaxies at $ z\sim3$ is 
$r_0\approx 4(6)h^{-1}$ Mpc for $\Omega_m=1.0 (0.2)$; note that this is
as large or larger than that of present--day galaxies. This fact
points out the danger in interpreting the evolution of galaxy correlation
functions from deep redshift surveys as being indicative of the overall
growth of structure by gravitational instability; samples of galaxies
at generally {\it smaller} redshifts have much smaller co--moving correlation
lengths (Carlberg et al 1997, Le F\'evre et al 1996). This probably results 
from the complicated and model--dependent interplay between effective
linear bias of halos in which galaxies have formed, and the (photometric) observational biases imposed
by the means by which the galaxies were selected, all of which are expected to
be redshift--dependent. 
In any case, the paradigm that galaxies form at high peaks in the density field
at early epochs that would be biased with respect to the  
overall mass distribution is strongly supported in our data, and
the observations are able to specify the spatial distribution of observable galaxies 
at early epochs, for comparison to models which include both the
dark matter component and recipes for star formation (e.g., Kauffman 1998;
Frenk 1998).
We are currently exploring more accurate statistics using our growing sample
that can be used to test various flavors of hierarchical models. At present,
combining information on the clustering properties {\it and} the observed abundance
of LBGs constrains the shape of the power-spectrum from galaxy to
cluster scales, and favors a universe with low overall matter density within the
context of CDM models (Adelberger et al 1998). It also suggests a very
tight relationship between dark matter halo mass and UV luminosity.   
We hope to have much more to say on such matters in the very near future. 

\section{Prospects for Galaxy Surveys at $z>4$}

In the process of obtaining the data for the $z\sim 3$ sample, we have recently
begun exploring the use of Lyman break color selection for a higher
redshift sample, with an expected median redshift $\langle z \rangle \sim 4.2$. 
In this case, the color selection is accomplished simply by adding an observed
$i$ band ($\lambda_{\rm eff} = 8100$ \AA) for those fields in which we already
have deep $G$ and ${\cal R}$ images. In order to make strong claims about
possible evolution of the star forming galaxy populations between $z \sim 3$ and
$z \sim 4.2$, it will be essential to establish an accurate redshift selection function.
Our initial forays into the spectroscopy of such candidates have been very instructive.
First, it certainly does work to use similar color--selection for this higher
redshift range (at the time of this writing we have confirmed 10 redshifts
in the range $3.9 \le z \le 4.5$), but it will be {\it far} more difficult
to obtain large galaxy samples at these redshifts as compared to $z \sim 3$. This is
partly due to a real decline in space density at a given luminosity (our preliminary
indications are that the abrupt decline beyond $z \sim 3$ as suggested in the
HDF [Madau et al. 1996] is supported by our ground--based data over much larger fields), partly due
to a slightly fainter apparent magnitude for a given absolute luminosity, but
mostly to the fact that the night sky background becomes much more troublesome
when the lines which one generally uses to secure the redshifts appear much farther
to the red. One of the consequences is that galaxies without strong Lyman $\alpha$ emission 
lines are exceedingly difficult to identify securely.  

\section{The Future}

It is clear that galaxies at very early epochs (the first
10\% of the age of the universe) are now open to the kinds of
wholesale, statistical studies that will provide significant
constraints on models of galaxy and structure formation. 
Issues that have clearly arisen at this meeting as being
very important are 1) to what extent is dust altering our
view of star formation in early galaxies, and therefore
our census of the global star formation history of the
universe? 2) What are the
clustering properties of early galaxies really telling us
about cosmological world models and/or how the galaxy
formation process works? 3) What are the present day
descendents of star forming galaxies seen at very high
redshift, and more generally how do objects observed
at one cosmic epoch relate to those observed at another? 
4) How are dark matter halos and observable galaxies related
to one another? 
The rapid developments on both observational and theoretical fronts,
many of which have been discussed at this meeting, 
will almost certainly lead to a great deal of progress on these
and other fundamental questions in just the next few years. 


%
%

\end{document}